\documentclass{svproc}

\usepackage[numbers]{natbib}
\usepackage{graphicx}
\usepackage{slashed}
\usepackage{subcaption}
\usepackage{amsmath}
\usepackage{amsfonts}
\usepackage{amssymb}
\usepackage{url}
\usepackage{amsbsy}
\usepackage{mathtools}

\usepackage[colorlinks]{hyperref}
\usepackage[usenames,dvipsnames]{color}

\usepackage[applemac]{inputenc}
\usepackage[T1]{fontenc}

\def\beq{\begin{equation}}
\def\eeq{\end{equation}}
\def\bea{\begin{eqnarray}}
\def\eea{\end{eqnarray}}
\def\be{\begin{equation}}
\def\ee{\end{equation}}
\def\ba{\begin{array}}
\def\ea{\end{array}}

\def\part{\partial}

\def\evar{\varepsilon}
\def\vecp{\mathbf{p}}

\def\vze{{\bf s}}
\def\ze{s}

\def\be{\begin{equation}}
\def\ee{\end{equation}}
\def\bea{\begin{eqnarray}}
\def\eea{\end{eqnarray}}
\newcommand{\eV}{\text{eV}}

\usepackage{url}

\begin{document}
\mainmatter              
\title{\textit{PAAI} in the sky : towards a particulate mechanism for Dark Energy and concordant Dark Matter}
\titlerunning{\textsl{... PAAI in the sky}}  
%
\author{R. B. MacKenzie\inst{1} \and M. B. Paranjape \inst{1}
 \and
U. A. Yajnik\inst{2}}
\authorrunning{R. B. MacKenzie \textit{et al.}} 
%
\tocauthor{R. B. MacKenzie,  M. B. Paranjape,  U. A. Yajnik}
\institute{Groupe de physique des particules, D\'epartement de
physique, Universit\'e de Montr\'eal, 
Montr\'eal, Qu\'ebec, CANADA,\\
\and
Physics Department, Indian Institute of Technology Bombay, Mumbai, India \\
\email{yajnik@iitb.ac.in}}

\maketitle              


\begin{abstract}
We propose the origins of Dark Energy in a hidden sector with a pair of very light fermions, oppositely charged under
an abelian gauge force $U(1)_X$ but of unequal mass. The system is dubbed PAAI, plasma 
which is abelian, asymmetric and idealised. For a range of the hidden fine structure constant values and the 
value of mass of the lightest fermion 
the PAAI is argued to simulate Dark Energy.  
Additional fermions from the same sector are shown to account for Dark Matter. Further, residual 
$X$-magnetic fields can mix with Maxwell electromagnetism to provides the seed  for cosmic-scale magnetic fields.  
Thus the scenario can explain several cosmological puzzles from within the same hidden sector\footnote{\footnotesize Prepared for \textit{Workshop on Frontiers in High
Energy Physics 2019}, A. Giri and R. Mohanta (eds.), Springer Proceedings in Physics 248, 2020}.
\end{abstract}

\section{Introduction}
There are several important unresolved issues in our current understanding of cosmology. Paramount among these are the problems 
of Dark Matter (DM) and Dark Energy (DE). Within the $\Lambda$-CDM model DM assists in galaxy formation and should be a  gas of non-relativistic particles, while the issue of DE is closely tied to that of the cosmological constant \cite{1989-Weinberg-Rev.Mod.Phys.},
since data \cite{2018-Aghanim.others-b} suggests that its energy density is constant over the epochs scanned by the cosmic 
microwave background (CMB).  
If treated as a dynamical phenomenon, DE demands an explanation for the equation of state $p=-\rho$  
in terms of relativistic  phenomena. From the point of view of naturalness, explaining a value of a dynamically generated quantity 
which is many orders of magnitude away from any of the scales of elementary particle physics or gravity is a major challenge. 
There are explanations that obtain such a  sector as directly related to  and derived from more powerful 
principles applicable at high scales  \cite{2011-Li.etal-Commun.Theor.Phys.} \cite{2018-Dey.etal-Nucl.Phys.} \cite{2004-Kapusta-Phys.Rev.Lett.}. On the other hand, extended and space filling objects, specifically domain walls as possible solutions to understanding 
Dark Energy have  been proposed earlier in a variety of scenarios \cite{Battye:1999eq,Battye:2007aa}\cite{Conversi:2004pi}\cite{Friedland:2002qs} \cite{Yajnik:2014eqa}. In this paper we pursue the latter approach, of  invoking new species of particles and their interactions at 
the new low mass scale, agnostic of their connection to the known physics other than gravity. 
A more extensive discussion of the results reported here can be found in \cite{MacKenzie:2019xth}.

We consider a new sector of particles with interaction mediated
by an unbroken abelian gauge symmetry denoted $U(1)_{X}$. The core of our mechanism involves the existence of
a fermionic species that  enters into a ferromagnetic state. As we will show, it is required to have an extremely small mass
and hence an extremely large magnetic moment; we dub this species the \textsl{magnino}\footnote{The term \textit{magnino} 
was earlier introduced in a different connotation in 
\cite{1988-Raby.West-Phys.Lett.a}\cite{1987-Raby.West-Phys.Lett.}}, denoted $M$. We assume 
that the medium remains neutral under the $X$-charge due to the presence of a significantly heavier 
species $Y$ of opposite charge which does not enter the 
collective ferromagnetic state. The wall complex resulting from the formation of magnetic domains then 
remains mutually bound, and due to interaction strength much larger than cosmic gravity, remains frozen.
The binding of the heavier species to this complex due to the requirement of $X$-electrical neutrality
then ensures that these particles remain unevolving, and after averaging over the large scales of the 
cosmic horizon act like a homogeneous space filling medium of constant density.

It is possible to explain DM within the same sector, including possible dark atoms formed by such species \cite{2009-Feng.etal-JCAP}\cite{2016-Boddy.etal-Phys.Rev.}
\cite{2014-Cline.etal-Phys.Rev.b,2012-Cline.etal-Phys.Rev.}.
This would also solve the \textit{concordance problem}, that is, the comparable energy densities carried in the cosmological
energy budget by the otherwise-unrelated components, DM and DE.
Further, the $X$-electromagnetism is expected to mix kinetically with the standard electromagnetism. The existence of cosmic magnetic fields at galactic and intergalactic scales 
\cite{Kulsrud:2007an}\cite{Durrer:2013pga}\cite{Subramanian:2015lua} is an outstanding puzzle of cosmology. 
Our mechanism relying as it does on spontaneous formation of domains of $X$-ferromagnetism has the potential to
provide the seeds needed to generate the observed fields through such mixing. 

In the following, in section \ref{sec:negpress}  we motivate the origin of negative pressure for extended objects in cosmology. 
In \ref{sec:PAAI} we discuss the calculation of the exchange energy for a spin polarised PAAI. Thus we motivate 
the possibility of occurrence of an extended structure of domain walls, and their metastable yet long lived nature.  
In section \ref{sec:singlemagnino} we discuss the main results of our proposal, obtaining suggestive values for the masses and abundances
for the scenario to successfully explain DE.
In section \ref{sec:flavo} we take up the possibility of concordant models with DM species arising compatible with this DE proposal. In section \ref{sec:cosmagfields} we discuss the possibility of obtaining an explanation of origin of cosmic magnetic fields from 
mixing of this $U(1)_X$ with standard electromagnetism.  Sec. \ref{sec:conclusion} contains the conclusion.

\section{Cosmic relics and the origin of negative pressure}
\label{sec:negpress}
A homogeneous, isotropic universe  is described by the Friedmann equation for the scale factor $a(t)$ 
supplemented by an equation of state relation $p=w\rho$. Extended relativistic objects in gauge theories in the 
cosmological setting\cite{Kibble:1980mv} are known to lead to negative values for $w$ \cite{KolTur,Dodelson}. 
A heuristic argument runs as follows. In the case of a 
frozen-out vortex line network, the average separation between string segments scales as $1/a^3$ but there is also an increment in 
the energy proportional to $a$ due to an average length of vortex network proportional to $a$ entering the physical volume. 
As such, the energy density of the  network has to be taken to scale as $1/a^2$, and we get the effective value $w=-1/3$. 
Likewise, for a domain wall complex, the effective energy density scales as $1/a$ and $w=-2/3$. By extension, for a relativistic
substance filling up space homogeneously, the energy density is independent of the scale factor, and has $w=-1$. 
In quantum theory this arises naturally as the vacuum expectation value of a relativistic scalar field. 
In the following, we consider a scenario that gives rise to a complex of domain walls whose separation scale 
is extremely small compared to the  causal horizon and  which remains fixed during expansion, and hence simulates 
an equation  of state $p=-\rho$.

\section{Ferromagnetic instability of PAAI}
\label{sec:PAAI}
A system of fermions  can be treated as a gas  of weakly interacting quasi-particles in the presence of oppositely charged much heavier ions 
or protons which are mostly spectators and serve to keep  the medium neutral. The  total energy of 
such a system can be treated as a functional of electron number 
density, according to the Hohenberg-Kohn theorem.  
In a relativistic setting, it becomes a functional of the covariant 4-current, and hence also of 
the electron spin density \cite{Rajagopal:1973}. In the Landau fermi liquid formalism the quasi-particle energy receives a correction from an interaction strength $f$
with other quasi-particles which can be determined from the forward scattering
amplitude $\mathcal{M}$  \cite{Baym:1975va}
\beq
f(\vecp\vze,\, 
\vecp'\vze')=\frac{m}{\evar^0(\vecp)}\frac{m}{\evar^0(\vecp')}\mathcal{M}(\vecp\ze,\, 
\vecp'\ze'),
\label{eq:fM}
\eeq
where $\evar^0$ is the free particle energy and $\mathcal{M}$ is the Lorentz-covariant $2\rightarrow2$ scattering amplitude in
a specific limit not discussed here. The exchange energy can equivalently be seen to arise as a 
two-loop correction to the self-energy of the fermion \cite{Chin:1977iz}. 
Using $f$ this one can compute the exchange energy $E_\text{xc}$, as
\beq
E_\text{xc} = \sum_{\pm \vze}\sum_{\pm \vze'}\int \frac{d^3 p}{(2\pi)^3}\frac{d^3 p'}{(2\pi)^3}
f(\vecp\vze,\, \vecp'\vze')n(\vecp,\vze)n(\vecp',\vze')
\eeq
and the effective quasi-particle energy is the kinetic energy of the quasi-particles with  renormalised mass parameter 
$E_{\text{kin}}$ plus the spin-dependent exchange energy in a spin-polarised background. For this purpose it 
is necessary to calculate  the self energy with a Feynman propagator in the presence of non-zero number density, and 
spin imbalance \cite{Bu:1984}. 

To set up a spin-asymmetric state, we introduce a parameter $\zeta$ such that the 
net density $n$ splits up into densities of spin up and down fermions as
\beq
n_{\uparrow}=n(1+\zeta) \quad \mathrm{and}\quad n_{\downarrow}=n(1-\zeta) 
\eeq
Correspondingly, we have Fermi momenta $p_{F\uparrow}=p_F(1+\zeta)^{1/3}$ and $p_{F\downarrow}=p_F(1-\zeta)^{1/3}$, with 
$p_F^3=3\pi^2n$. The exchange energy was calculated in \cite{Bu:1984} and the final expression is too long to be
quoted in this presentation. However the leading order expansions in $\beta=p_F/m$ for the fully polarised case 
$\zeta=1$  is\cite{MacKenzie:2019xth}
\bea
E_\text{kin}(\zeta=1)
&=&m^4\left\lbrace \frac{\tilde{\beta}^5}{20 \pi ^2}-\frac{\tilde{\beta}^7}{112 \pi ^2}
+O\left(\beta^{9}\right) \right\rbrace \\
E_\text{xc}(\zeta=1) &=&
-\alpha_{X} m^4 \left\lbrace \frac{{\tilde{\beta}}^4}{2 \pi ^2} - \frac{7 {\tilde{\beta}}^6}{27 \pi ^2}
+O\left(\tilde{\beta}^{8}\right)  \right\rbrace
\eea
where $\tilde{\beta}=2^{1/3}\beta$. The $\zeta=0$ case has same leading power laws with different coefficients.
Thus the exchange energy tends to lower the quasi-particle energy parametrically determined by $\alpha$, with 
either $\zeta=0$ or $\zeta=1$ becoming the absolute minimum depending on $\beta$. For comparison, in this notation, the rest mass
energy of the degenerate gas is $E_{\text{rest}}=m^4 \beta^3/(3\pi^2)$.

\begin{figure}[htb]
 \centering{\includegraphics[width=.8\linewidth]{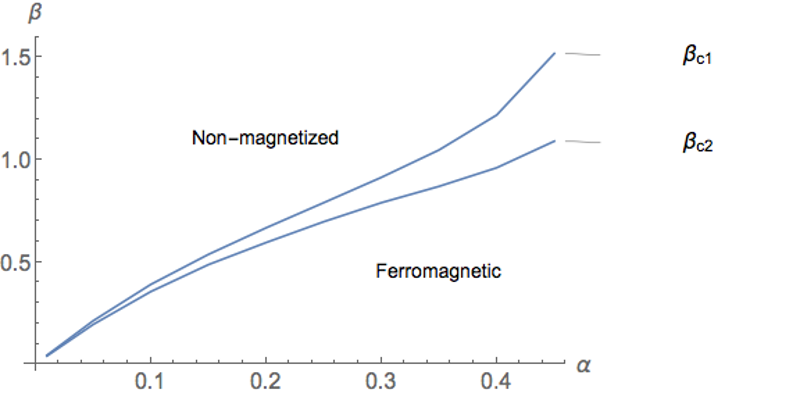}}
  \caption{Phase plot in the fine structure constant $\alpha$ vs $\beta=p_F/m$ plane 
showing the   allowed region of spontaneous ferromagnetism}
  \label{fig:phasedia}
\end{figure}

Exploring the energy expression presents three possibilities; $\zeta=1$ is not a minimum at all, $\zeta=1$ is 
a local minimum but $E(0)<E(1)$ i.e. a metastable vacuum and finally, $\zeta=1$ is the absolute minimum with $\zeta=0$ unstable vacuum.
In Fig.~\ref{fig:phasedia} we have plotted the approximate regions of the three phases in the parameter space.

\subsection{Evolution and stability of domain walls}
\label{sec:dw}
We expect domain walls to occur in this spin polarised medium just like in ferromagnets. However due to the 
$SU(2)$ of spin being simply connected, the defects are not topologically stable and can unwind. 
However these processes are suppressed by a competition between the gradient energy and the
extra energy stored in the domain walls, and there is a Ginzberg temperature $T_G$ \cite{Kibble:1980mv} below which thermal fluctuations
cannot destabilise the walls trivially.  The mechanism for destabilisation is then the one studied in detail in \cite{Preskill:1992ck}.
The rate for such decay is governed by an exponential factor $\exp(-B/\lambda)$
\cite{Kobzarev:1974cp} where the exponent is the Euclidean action of a suitable "bounce" solution
connecting the false and the true vacua \cite{Coleman:1977py}.
On phenomenological grounds we need this complex to be stable for $\approx 10^{17}sec$. 
The bounce $B$ is typically $\propto 1/\lambda$  where $\lambda$ is a generic dimensionless coupling constant.
Then large suppression factors $\sim10^{-30}$ are natural for $\lambda \sim 0.01$. The other mechanism for 
disintegration of the DW network resides in the magnino gas  becoming non-degenerate.

\section{A minimal model for Dark Energy}
\label{sec:singlemagnino}
We consider a hitherto unobserved sector with particle species we generically call $M$ and
$Y$. They are assumed to be oppositely charged under a local abelian group $U(1)_{X}$ with fine structure constant $\alpha_{X}$.
The mass $m_M$ of $M$ is assumed in the sub-eV range while the $Y$ mass $m_Y$ is assumed to be much larger. 
Charge neutrality requires that
the number densities of the two species have to be equal, in turn this means that the Fermi energies
are also the same. The hypothesis of larger mass is to ensures that $Y$ with Compton wavelength $M^{-1}\ll p_F^{-1}$
does not enter into a collective magnetic phase.

We  start our considerations at time $t_1$  when the temperature is just below
$T_G$ so that the wall complex has materialised. The parameters of this wall complex are $\omega$, the thickness of individual 
walls and $L$ the average separation between walls. On the scale of
the horizon, the wall complex behaves just like a space filling homogeneous substance. Further, due to the demand 
of neutrality, the heavier gas $Y$ cannot
expand either, although it has no condensation effects. Let us denote the number density of the magninos
trapped in the walls to be $n^X_{\mathrm{walls}}$ and the remainder residing in the enclosed domains
by $n^X_{\mathrm{bulk}}$. Averaged (coarse grained) over a volume much larger than the $L^3$, this gives the
average number density of the magninos to be
\be
\langle n^X \rangle = \frac{\omega}{L} n^X_{\mathrm{walls}} + \left( 1- \frac{\omega}{L} \right) n^X_{\mathrm{bulk}}
\ee 
And from the neutrality condition we have
\be
\langle n^X \rangle = \langle n^Y \rangle
\ee
Then we can demand that PAAI in this phase acts as the DE, so that assuming $Y$ to be non-relativistic, 
and ignoring other contributions,
\be
\rho^Y \approx m_Y \langle n^Y \rangle = \rho_{\text{DE}} = 2.81\times10^{-11} (\eV)^4
\ee
We can express the number density of $Y$ as a ratio of the number density $n_\gamma=3.12\times 10^{-12}(\eV)^3$ 
of photons, and set $\eta^Y= \langle n^Y \rangle/n_\gamma$. Then we can obtain conditions that determine the ratio
\be
\label{eq:SImupperbound}
\frac{m_M}{m_Y}=\frac{\beta^Y}{\beta}\approx (\eta^Y)^{4/3} \times 10^{-6} \ll 1
\ee
These are the essential constraints determining the key parameters of our model. 
Then we find that $m_M$ ranges over $10^{-4}$ to $10^{-6}$ eV corresponding to $\eta^Y$ ranging 
from  $10^{-5}$ to $10^{-8}$; and $m_Y$ respectively ranges from $1$keV to $1$GeV. 
The details can be found in \cite{MacKenzie:2019xth}.

\section{Flavoured models and cosmic concordance}
\label{sec:flavo}
It is now interesting to explore whether this hidden sector admitting $X$-ferro\-magnetic condensation mechanism 
also has possibilities for the  DM. This requires the existence of additional number of stable species which become 
non-relativistic while  the lightest particles continue to simulate Dark Energy. Let us denote the \textsl{general} 
requirements to be obeyed by such \textsl{flavoured} scenarios to be \textbf{GF}.
The wish list of such requirements is
\begin{description}
\item[GF1] The charges of these species under $U(1)_{X}$ are opposite in sign for $M$-type versus $Y$-type. 
However we leave open the possibility that the magnitudes of these charges can be small integer multiples of each other.
\item[GF2] The  heavier species of $M$-types and $Y$-types should be stable against decay into the corresponding 
lighter ones even if their $Q_{X}$ charges tally. This is analogous to flavour symmetry in the observed 
sector, where the purely electromagnetic  conversion of heavier leptonic  flavours into lighter ones is not observed.   
\item[GF3] The lightest pair $M_1$ and $Y_1$ (more generally at least one effective degree of freedom of species of each type) 
have  equal and opposite charges, and satisfy the requirement of the DE scenario of Sec. \ref{sec:singlemagnino}.
\item[GF4] The heavier species (more generally the remainder degrees of freedom) do not undergo condensation.
\end{description}

Within these general criteria the simplest scenario that can be thought of may be called \textbf{FI}. It has the 
following straightforward requirements 
\begin{description}
\item[FI-1] The pair of species $M_2$ and $Y_2$ with $Q_{X}(M_2)=-Q_{X}(Y_2)$
\item[FI-2] This pair of species accounts for the observed DM.
\end{description}
Thus we demand, with $n_{M2}=n_{Y2}$ designating the number densities, that
\be
(m_{M2}+m_{Y2})n_{Y2} = \rho_\text{DM}= 1.04\times 10^{-11} (\eV)^4
\ee
so that  
\be
m_{M2}+m_{Y2}=\left( \frac{1}{\eta^{Y2}}\right) 3.33 \eV
\label{eq:DMconstraint}
\ee
In order for either of  $X_2$ or $Y_2$, or both together to act as DM, the right hand
side of the above equation has to be at least a few keV to satisfy the generally accepted phenomenological 
requirement on Dark Matter. Thus we need 
\be
\eta^{Y2}\lesssim 10^{-3} \qquad \text{to ensure DM mass}\gtrsim \text{keV}
\label{eq:etaYconstraint}
\ee
From Sec.~\ref{sec:singlemagnino}, we have that $\eta^Y$ can take on any value $\lesssim 10^3$ and account for
DE adequately. The DM constraint on the second flavour restricts its abundance to $\lesssim 10^{-3}$.
In this scenario $\eta^Y$ and $\eta^{Y2}$ need not be related, and a few orders of magnitude difference in abundance 
could be easily explained by dynamics occurring within that sector in an expanding universe.
Further we shall see later that the large value of $\eta^Y$ makes the scenario capable of explaining 
the origins of cosmic magnetic fields, while the small $\eta^{Y2}$ value can separately solve the DM puzzle. 

The scenario $FI$ requires that at least one of $M_2$ and $Y_2$ is heavy enough to be the DM particle.
But it leaves the mass of the other particle undetermined. A scenario that is more restrictive about the mass of $M_2$ could arise 
as follows, and we denote this scenario \textbf{FII}. 
\begin{description}
\item[FII-1] There are two species $M_1$ and $M_2$, of the same charge $Q_{X}(M_2)=Q_{X}(M_1)$.
\item[FII-2] $\eta^{M1}= \sigma \eta^{M2}$  where $\sigma$ is a numerical factor
\item[FII-3] Only $M_1$ is the magnino, capable of condensing.
\item[FII-4] There is only one species of $Y$ type, with $Q_{X}(Y)=-Q_{X}(M_1)$.
\end{description}
For neutrality of the medium we need $\eta^Y=\eta^{M1}+\eta^{M2}$. Then in this scenario, the fraction equivalent to
$\eta^{M1}$  of the $Y$ particles will suffice to keep the condensed state of $M_1$ neutral, and thus the mass of $Y$ will 
be determined as in Sec.~\ref{sec:singlemagnino}
The remainder $Y$ particles, in abundance $\eta^{M2}$ scale like free matter particles.
Then analogous to conditions Eq.s \eqref{eq:DMconstraint} \eqref{eq:etaYconstraint}, we get
\bea
m_{M2}+m_{Y}&=&\left( \frac{1}{\eta^{Y2}}\right) 3.33 \eV \\
\eta^{M2}&\lesssim& 10^{-3}
\eea
The point is that $m_{Y}$ is already determined by the value of $\eta^{M1}$ from DE Condition, and if 
$\eta^{M1}\gtrsim 1$  then mass of $Y$ would be determiend to be too small to be DM candidate.
In this case, without proliferating unknown mass values, $m_{M2}$ can be the DM candidate.

This Dark Matter sector is along the lines of \cite{2009-Feng.etal-JCAP}, and through out its history could have been
partially ionised and could be progressively becoming neutral. In particular it represents the class of self interacting
Dark Matter including van der Waals forces that may result between such atoms due to very low binding energy.  
It has  been argued for example in  \cite{2017-Kamada.etal-Phys.Rev.Lett.} that such a model potentially explains
the diversity in the rotation curves of galaxies.

\section{Origin of cosmic magnetic fields}
\label{sec:cosmagfields}
The origin and evolution of galactic scale magnetic fields is an open
question \cite{Kulsrud:1999bg,Kulsrud:2007an}. In particular the extent of seed magnetic 
field as against that generated by subsequent motion is probably 
experimentally distinguishable \cite{Durrer:2013pga}\cite{Subramanian:2015lua}. 
In the present case, we can estimate the field strength of the $X$-magnetism in each domain, and since the 
domain structure is completely random we expect zero large scale magnetic field on the average. 
Residual departure from this average can be estimated by assuming that the deviation from the mean grows as $\sqrt{N}$ as
we include $N$ domains. Thus if the $X$-magnetic field in individual domains has the value $B_\text{dom}$ then on the scale of 
galactic clusters  $L_\text{gal}$ it possesses a root mean square value  $\overline{\Delta B}\equiv B_\text{dom}(L/L_\text{gal})^{3/2}$. 

Assuming $U(1)_{X}$ field mixes kinetically with standard electromagnetism thr\-ough term of the form 
$\xi F^{\mu \nu}F^X_{\mu \nu}$,  we consider the possibility of a 
seed of $10^{-30}$T with a coherence length of $0.1$ kpc$\sim3\times 10^{18}$ meter obtained with $\xi=10^{-8}$. With $B_\text{dom}$
calculated from the formalism of Sec. \ref{sec:PAAI} we can obtain
\be
\overline{\Delta B}_{\mathrm{seed}}=10^{-30}T\sim 10^{-8}\times\left(\frac{m_M}{\eV}\right)^2\left( \frac{\alpha_X}{\alpha}\right)^{1/2} \beta^3 
\left( \frac{L}{\mathrm{meter}}  \right)^{3/2}\times 10^{-40} T 
\ee
From this, representative values for $L$ for $\beta=0.1$ are  in the range  $10^{14}$-$10^{15}$ meter which is a fraction of 
the Milky Way size.  A detailed treatment to estimate the residual
fluxes on large coherence length scales could trace the  statistics of flux values in near neighbour domains and the rate at which 
the magnetic flux could undergo percolation, providing perhaps a smaller value for $L$, comparable to the above estimate.

\section{Conclusions}
\label{sec:conclusion}
We have proposed the possibility of a negative pressure medium as arising from 
nothing more radical than a peculiar ground state of a pair of asymmetric fermion species interacting through an
unbroken abelian  gauge force.  In an attempt to highlight the potential utility of the PAAI to cosmology, 
specifically to DE and to cosmic ferromagnetism, we have been agnostic about the earlier history of this sector. 
A study of temperature dependence of this phenomenon as also phenomenological inputs from the cosmic dawn data would help to 
sharpen this scenario. Large scale magnetic fields could arise from the same scenario and Dark Matter can
be accommodated within the same hidden sector.

\section{ACKNOWLEDGEMENTS}
We thank NSERC, Canada for financial support and the Ministère des relations internationales et la 
francophonie of the Government of Québec for financing within the cadre of the Québec-Maharashtra 
exchange.  RBM and MBP also thank IIT Bombay for financial support and hospitality.


\bibliographystyle{natbib}

\end{document}